\documentclass[11pt,english,reqno]{amsart}
\usepackage[T1]{fontenc}
\usepackage{mathrsfs}
\usepackage{amsthm}
\usepackage{amsbsy}
\usepackage{amstext}
\usepackage{amssymb}
\usepackage{mathdots}

\makeatletter
\numberwithin{equation}{section}
\numberwithin{figure}{section}
\theoremstyle{plain}
\newtheorem{thm}{\protect\theoremname}[section]
  \theoremstyle{remark}
  \newtheorem{rem}[thm]{\protect\remarkname}

\usepackage{graphicx}
\graphicspath{{graphics//}}
\usepackage{color}
\newcommand{\ie}{\textit{i.e.}}
\newcommand{\eg}{\textit{e.g.}}
\usepackage{amscd}
\usepackage{hyperref}
\numberwithin{equation}{section} 

\makeatother

\usepackage{babel}
  \providecommand{\remarkname}{Remark}
\providecommand{\theoremname}{Theorem}

\begin{document}

\global\long\def\ga{\alpha}
\global\long\def\gb{\beta}
\global\long\def\ggm{\gamma}
\global\long\def\go{\omega}
\global\long\def\ge{\epsilon}
\global\long\def\gs{\sigma}
\global\long\def\gd{\delta}
\global\long\def\gD{\Delta}
\global\long\def\vph{\varphi}
\global\long\def\gf{\varphi}
\global\long\def\gk{\kappa}

\global\long\def\wh#1{\widehat{#1}}
\global\long\def\bv#1{\mathbf{#1}}
\global\long\def\bs#1{\boldsymbol{#1}}

\global\long\def\ui{\wh{\boldsymbol{\imath}}}
\global\long\def\uj{\wh{\boldsymbol{\jmath}}}
\global\long\def\uk{\wh{\textbf{\em k}}}

\global\long\def\bosy#1{\boldsymbol{#1}}

\global\long\def\vect#1{\overline{\mathbf{#1}}}

\global\long\def\uI{\widehat{\mathbf{I}}}
\global\long\def\uJ{\widehat{\mathbf{J}}}
\global\long\def\uK{\widehat{\mathbf{K}}}

\global\long\def\uv#1{\widehat{\mathbf{#1}}}

\global\long\def\cross{\boldsymbol{\times}}

\global\long\def\ddt{\frac{\dee}{\dee t}}
\global\long\def\dbyd#1{\frac{\dee}{\dee#1}}

\global\long\def\fall{,\quad\text{for all}\quad}

\global\long\def\reals{\mathbb{R}}

\global\long\def\rthree{\reals^{3}}
\global\long\def\rsix{\reals^{6}}

\global\long\def\les{\leqslant}

\global\long\def\ges{\geqslant}

\global\long\def\dee{\mathrm{\mathrm{d}}}

\global\long\def\from{\colon}

\global\long\def\tto{\longrightarrow}

\global\long\def\abs#1{\left|#1\right|}

\global\long\def\isom{\cong}

\global\long\def\comp{\circ}

\global\long\def\cl#1{\overline{#1}}

\global\long\def\fun{\varphi}

\global\long\def\interior{\mathrm{Int\,}}
\global\long\def\diver{\mathrm{div\,}}

\global\long\def\sign{\mathrm{sign\,}}

\global\long\def\dimension{\mathrm{dim\,}}

\global\long\def\esssup{\mathrm{ess}\,\sup}

\global\long\def\ess{\mathrm{{ess}}}

\global\long\def\kernel{\text{Kernel}\,}

\global\long\def\support{\mathrm{Supp}\,}

\global\long\def\image{\mathrm{Image\,}}
\global\long\def\resto#1{|_{#1}}

\global\long\def\incl{\iota}

\global\long\def\rest{\rho}
\global\long\def\extnd{e_{0}}

\global\long\def\proj{\pi}

\global\long\def\sphere{S^{2}}
\global\long\def\hemis{H}

\global\long\def\ino#1{\int\limits _{#1}}

\global\long\def\half{\frac{1}{2}}

\global\long\def\shalf{{\scriptstyle \half}}

\global\long\def\third{\frac{1}{3}}

\global\long\def\empt{\varnothing}

\global\long\def\paren#1{\left(#1\right)}

\global\long\def\bigp#1{\bigl(#1\bigr)}

\global\long\def\biggp#1{\biggl(#1\biggr)}

\global\long\def\Bigp#1{\Bigl(#1\Bigr)}

\global\long\def\braces#1{\left\{  #1\right\}  }

\global\long\def\sqbr#1{\left[#1\right]}

\global\long\def\norm#1{\|#1\|}

\global\long\def\trps{^{\mathsf{T}}}

\global\long\def\contr{\raisebox{0.4pt}{\mbox{\ensuremath{\lrcorner}}}\,}

\global\long\def\pis{x}

\global\long\def\pib{X}

\global\long\def\body{B}

\global\long\def\bdry{\partial}

\global\long\def\gO{\varOmega}

\global\long\def\reg{R}

\global\long\def\bdom{\bdry\reg}

\global\long\def\bndo{\partial\gO}
\global\long\def\pbndo{\Gamma}
\global\long\def\bndoo{\pbndo_{0}}
 \global\long\def\bndot{\pbndo_{t}}

\global\long\def\cloo{\cl{\gO}}

\global\long\def\nor{\boldsymbol{n}}
\global\long\def\nora{\nor}
\global\long\def\norb{\boldsymbol{u}}
\global\long\def\norc{v}

\global\long\def\dA{\,\dee A}

\global\long\def\dV{\,\dee V}

\global\long\def\eps{\varepsilon}

\global\long\def\vs{\mathbf{W}}
\global\long\def\avs{\mathbf{V}}

\global\long\def\vbase{\boldsymbol{e}}

\global\long\def\vf{w}

\global\long\def\avf{u}

\global\long\def\stn{\varepsilon}

\global\long\def\rig{r}

\global\long\def\rigs{\mathcal{R}}

\global\long\def\qrigs{\!/\!\rigs}

\global\long\def\qd{\!/\,\!\kernel\diffop}

\global\long\def\dis{\chi}

\global\long\def\fc{F}

\global\long\def\st{\sigma}

\global\long\def\bfc{b}

\global\long\def\sfc{t}

\global\long\def\stm{S}

\global\long\def\sts{\varSigma}

\global\long\def\ebdfc{T}
\global\long\def\optimum{\st^{\mathrm{opt}}}
\global\long\def\scf{K}

\global\long\def\cee#1{C^{#1}}

\global\long\def\lone{L^{1}}

\global\long\def\linf{L^{\infty}}

\global\long\def\lp#1{L^{#1}}

\global\long\def\ofbdo{(\bndo)}

\global\long\def\ofclo{(\cloo)}

\global\long\def\vono{(\gO,\rthree)}

\global\long\def\vonbdo{(\bndo,\rthree)}
\global\long\def\vonbdoo{(\bndoo,\rthree)}
\global\long\def\vonbdot{(\bndot,\rthree)}

\global\long\def\vonclo{(\cl{\gO},\rthree)}

\global\long\def\strono{(\gO,\reals^{6})}

\global\long\def\sob{W_{1}^{1}}

\global\long\def\sobb{\sob(\gO,\rthree)}

\global\long\def\lob{\lone(\gO,\rthree)}

\global\long\def\lib{\linf(\gO,\reals^{12})}

\global\long\def\ofO{(\gO)}

\global\long\def\oneo{{1,\gO}}
\global\long\def\onebdo{{1,\bndo}}
\global\long\def\info{{\infty,\gO}}

\global\long\def\infclo{{\infty,\cloo}}

\global\long\def\infbdo{{\infty,\bndo}}

\global\long\def\ld{LD}

\global\long\def\ldo{\ld\ofO}
\global\long\def\ldoo{\ldo_{0}}

\global\long\def\trace{\gamma}

\global\long\def\pr{\proj_{\rigs}}

\global\long\def\pq{\proj}

\global\long\def\qr{\,/\,\reals}

\global\long\def\aro{S_{1}}
\global\long\def\art{S_{2}}

\global\long\def\mo{m_{1}}
\global\long\def\mt{m_{2}}

\global\long\def\yieldc{B}

\global\long\def\yieldf{Y}

\global\long\def\trpr{\pi_{P}}

\global\long\def\devpr{\pi_{\devsp}}

\global\long\def\prsp{P}

\global\long\def\devsp{D}

\global\long\def\ynorm#1{\|#1\|_{\yieldf}}

\global\long\def\colls{\Psi}

\global\long\def\ssx{S}

\global\long\def\smap{s}

\global\long\def\smat{\chi}

\global\long\def\sx{e}

\global\long\def\snode{P}

\global\long\def\elem{e}

\global\long\def\nel{L}

\global\long\def\el{l}

\global\long\def\ipln{\phi}

\global\long\def\ndof{D}

\global\long\def\dof{d}

\global\long\def\nldof{N}

\global\long\def\ldof{n}

\global\long\def\lvf{\chi}

\global\long\def\lfc{\varphi}

\global\long\def\amat{A}

\global\long\def\snomat{E}

\global\long\def\femat{E}

\global\long\def\tmat{T}

\global\long\def\fvec{f}

\global\long\def\snsp{\mathcal{S}}

\global\long\def\slnsp{\Phi}

\global\long\def\ro{r_{1}}

\global\long\def\rtwo{r_{2}}

\global\long\def\rth{r_{3}}

\global\long\def\mind{\alpha}
\global\long\def\vb{\xi}

\global\long\def\vbt{E}
\global\long\def\fib{\mathbf{V}}

\global\long\def\jetb#1{J^{#1}}

\global\long\def\jetm#1{j_{#1}}

\global\long\def\sobp#1#2{W_{#2}^{#1}}

\global\long\def\inner#1#2{\left\langle #1,#2\right\rangle }

\global\long\def\fields{\sobp pk(\vb)}

\global\long\def\bodyfields{\sobp p{k_{\partial}}(\vb)}

\global\long\def\forces{\sobp pk(\vb)^{*}}

\global\long\def\bfields{\sobp p{k_{\partial}}(\vb\resto{\bndo})}

\global\long\def\loadp{(\sfc,\bfc)}

\global\long\def\strains{\lp p(\jetb k(\vb))}

\global\long\def\stresses{\lp{p'}(\jetb k(\vb)^{*})}

\global\long\def\diffop{D}

\global\long\def\strainm{E}

\global\long\def\incomps{\vs_{\yieldf}}

\global\long\def\devs{L^{p'}(\eta_{1}^{*})}

\global\long\def\incompsns{L^{p}(\eta_{1})}

\global\long\def\dists{\mathcal{D}'}
\global\long\def\testfs{\mathcal{D}}

\global\long\def\prop{P}

\global\long\def\aprop{Q}

\global\long\def\flux{T}

\global\long\def\fform{\tau}

\global\long\def\dimn{n}

\global\long\def\sdim{{\dimn-1}}

\global\long\def\prodf{{\scriptstyle \smallsmile}}

\global\long\def\ptnl{\varphi}

\global\long\def\form{\omega}

\global\long\def\dens{\rho}

\global\long\def\simp{s}

\global\long\def\cell{C}

\global\long\def\chain{B}

\global\long\def\ach{A}

\global\long\def\coch{X}

\global\long\def\scale{s}

\global\long\def\fnorm#1{\abs{#1}^{\flat}}

\global\long\def\chains{\mathcal{A}}

\global\long\def\ivs{\boldsymbol{U}}

\global\long\def\mvs{\boldsymbol{V}}

\global\long\def\cvs{\boldsymbol{W}}

\global\long\def\subbs{\mathcal{B}}

\global\long\def\elements{\mathcal{E}}

\global\long\def\element{E}

\global\long\def\nodes{\mathcal{N}}

\global\long\def\node{N}

\global\long\def\psubbs{\mathcal{P}}

\global\long\def\psubb{P}

\global\long\def\matr{M}

\global\long\def\nodemap{\nu}

\global\long\def\prop{p}
\global\long\def\radi{K}
\global\long\def\radvec{\mathbf{i}}
\global\long\def\radint{i}
\global\long\def\mrad{\mathbf{I}}
\global\long\def\mradint{I}
\global\long\def\irrad{\phi}
\global\long\def\Hemis{\mathscr{H}}
\global\long\def\phasesp{\rthree\times\sphere}
\global\long\def\sterad{\omega}
\global\long\def\meters{w}
\global\long\def\coneu{\hat{u}}
\global\long\def\moment{M}
\global\long\def\pdists{U}
\global\long\def\prerad{\kappa}
\global\long\def\pmd{v}
\global\long\def\tradi{J}
\global\long\def\measrs{M}
\global\long\def\total{\Phi}
\global\long\def\mradi{J}

\title{Radiative Transfer and Flux Theory }

\author{Reuven Segev and Joe goddard}

\address{Department of Mechanical Engineering, Ben-Gurion University, Beer-Sheva,
Israel; Department of Mechanical and Aerospace Engineering, University
of California, San Diego, U.S.A.}



\dedicatory{\emph{}}

\date{\today}
\begin{abstract}
The fundamental notions of radiative transfer, \eg, Lambert's cosine
rule, are studied from the point of view of flux and stress theory
of continuum mechanics. For the classical case, where the radiance
is distributed regularly over the unit sphere, it is shown that Lambert's
rule follows from a balance law for the transfer of radiative power
in each direction $\norb$ of the sphere, together with the appropriate
Cauchy postulates and the additional assumption that the corresponding
flux vector field $\radvec_{\norb}$ be parallel to $\norb$. An analogous
theory is presented for the irregular case where the distribution
of radiance on the sphere is given as a Borel measure.

\end{abstract}
\maketitle

 
\thispagestyle{empty}

\noindent

\section{Introduction}

We study the basic elements of radiometry from the point of view of
flux and stress theory of continuum mechanics. Specifically, let $x$
be a point on the boundary $\bdom$ of some region $\reg$ in space,
let $\norb$ be some unit vector and let $\theta$ denote the angle
between $\norb$ and the unit normal $\nor$ to $\bdom$ at $x$.
Then, we recall that ignoring the dependence on the wavelength, the
radiative energy flux density, the irradiance, $E$, out of $\bdom$
at $x$ is traditionally given in terms of the radiance field (radiative
intensity) $I(x,\norb)$ by
\begin{equation}
E=\int_{\sphere}I(x,\norb)\cos\theta(\norb)\,\dee\sterad(\norb),
\end{equation}
where $\sphere$ is the unit sphere containing all directions $\norb$
and $\sterad$ denotes the solid angle on the sphere. The relation
\begin{equation}
\frac{\dee E}{\dee\sterad}=\mradint(x,\norb)\cos\theta
\end{equation}
is usually referred to as Lambert's cosine rule (\eg, \cite[p. 20]{Preisendorfer1965}\cite[p. 20]{Siegel1992}).
Thus, the total radiative energy flux out of $\reg$ is given by
\begin{equation}
P=\int_{\bdom}\left[\int_{\sphere}I(x,\norb)\cos\theta(\norb)\,\dee\sterad(\norb)\right]\dee A.\label{eq:TotPowerClass}
\end{equation}

We show below that these traditional relations and more general expressions
for the radiative energy flux follow from the theory of Cauchy fluxes
supplemented with a single additional postulate. For the case where
the radiance distribution at a point is a real valued function defined
on $\sphere$, Cauchy's postulates are applied to the flux of energy
in a generic direction $\norb$ to yield a flux vector field $\radvec_{\norb}$.
Alternatively, and this is the approach used later for the more general
situation where the distribution at a point may be as singular as
a measure, the collection of flux vector fields, $\{\radvec_{\norb}\}$,
$\norb\in\sphere$, may be viewed as an infinite dimensional stress-like
tensor field in the following sense. Given a point $x$ and the unit
normal vector $\nor$ as above, the traditional stress tensor determines
the 3-dimensional vector of traction on $\bdom$ at $x$. In analogy,
the infinite dimensional tensor corresponding to radiance determines
the distribution of flux on the sphere $\sphere$ at $x$, an element
of an infinite dimensional vector space $C^{0}(\sphere)$ of continuous
functions on the sphere. From a different point of view, the situation
is analogous to mixture theory where a flux vector field is associated
with each of the constituents. It is noted that the constituents of
the analogous mixture should be labeled by a continuous variable on
the sphere. Thus, the approach presented here offers a local point
of view to the apparent non-local character of radiometry. One should
also mention that a formal derivation of radiometry can be based on
the principles of electromagnetism (\eg, \cite{born,Gershun39}),
where the radiance is obtained as average of the Poynting vector.

The flux fields $\{\radvec_{\norb}\}$ have an additional property
that is characteristic of radiation: $\radvec_{\norb}$ is parallel
to $\norb$. This additional property implies the traditional law
of radiometry as presented above.

Replacing the space of continuous functions $C^{0}(\sphere)$ by the
space $M(\sphere)$ of Borel measures on the sphere, enables one to
consider irregular radiation distributions such as monodirectional
radiation. The property of radiation as described above cannot be
applied directly to irregular distributions of radiation on the sphere
because $\radvec_{\norb}$ is not well defined. However, an analogous
generalized formulation of the property is suggested by using some
basic notions of measure theory.

\section{Radiance Flux Fields: Classical Analysis}

\subsection{Traditional Cauchy Fluxes\label{sub:Traditional-Cauchy-Fluxes}}

Traditionally, Cauchy's flux theory considers an extensive scalar
property $\prop$ of regions $\reg$ in the physical space which is
represented here by $\rthree$. It is assumed that for each region
$\reg\subset\rthree$, there is a field $\psi_{\reg}:\bdom\to\reals$,
the flux density of $\prop$ out of $\reg$, that represents the total
flux $\Psi_{\reg}$ of $\prop$ through the boundary $\bdom$ in the
form
\begin{equation}
\Psi_{\reg}=\int_{\bdom}\psi_{\reg}\dA.
\end{equation}
A collection $\{\psi_{\reg}\}$ of fields on the boundaries of all
regions $\reg\subset\rthree$ will be referred to as a \emph{flux
system. }Cauchy's flux theory is developed on the basis of the following
assumptions.

\subsubsection*{Boundedness (Balance)}

Let $\abs{\reg}$ denote the volume of the region $\reg$. It is assumed
that there is a positive constant $C$ such that 
\begin{equation}
\abs{\Psi_{\reg}}\les C\abs{\reg}.\label{eq:Cauchy-Boundedness}
\end{equation}

The boundedness assumption above is usually motivated by a balance
principle. It is assumed that the total flux of the property $\prop$
out of $\reg$ is equal to the rate of production of the property
within $\reg$ minus the rate of change of the total of $p$ in $\reg$.
Thus, let $\rho$ be the density of $\prop$ in $\rthree$ and let
$s$ be the density of the source of $\prop$ in $\rthree$, then
\begin{equation}
\int_{\reg}\dot{\rho}\dV+\int_{\bdom}\psi_{\reg}\dA=\int_{\reg}s\dV.\label{eq:balanceClassical}
\end{equation}
The last balance equation, once supplemented by appropriate boundedness
assumptions for $\dot{\rho}$ and $s$, will imply (\ref{eq:Cauchy-Boundedness}). 

It is observed that this assumption rules out irregular sources of
the property such as those concentrated on surfaces, lines, or points.

\subsubsection*{Cauchy's postulate of locality}

Considering the dependence of the field $\psi_{\reg}$ on the region
$\reg$, Cauchy's locality postulate implies that this dependence
is of a very short range. Specifically, it states that for $x\in\bdom$,
$\psi_{\reg}(x)$ depends on $\reg$ only through the unit normal
$\nor(x)$ to $\bdom$ at $x$. Let $\sphere$ denote the 2-dimensional
unit sphere in $\rthree$ containing the collection of unit vectors.
It follows that there is a function $\tau:\rthree\times\sphere\to\reals$
such that 
\begin{equation}
\psi_{\reg}(x)=\tau(x,\nor(x)).
\end{equation}

\subsubsection*{Regularity}

It is assumed that $\tau$ is a smooth function.

Naturally, we will refer to a flux system satisfying these assumptions
as a \emph{Cauchy Flux System}.

\subsubsection*{Cauchy's Flux Theorem}

The boundedness and regularity assumptions above imply that the dependence
of $\tau$ on $\nor$ is linear. Since any linear function of $\nor$
may be represented by an inner product with $\nor$, it follows that
for a Cauchy flux system there is a vector field $\flux:\rthree\to\rthree$
such that 
\begin{equation}
\psi_{\reg}(x)=\tau(x,\nor(x))=\flux(x)\cdot\nor(x).\label{eq:ClassicalCauchy}
\end{equation}
We refer to $\flux$ as the \emph{flux vector field} associated with
the property $p$.

Consider for example the case where the property under consideration
is the radiation energy so that $\Psi_{\reg}$ is interpreted at the
total flux of radiation energy through the boundary of a region $\reg$.
Then, under the foregoing assumptions, Cauchy's flux theorem implies
that there is a vector field $\boldsymbol{q}$ such that the flux
of radiation energy out of the boundary of a region $\reg$ is given
by (\ref{eq:ClassicalCauchy}). In fact, for this case, $\bs q\cdot\nor$
is the irradiation as implied by equation (\ref{eq:TotPowerClass}).

\subsubsection*{The Differential Balance Equation}

Using Cauchy's theorem in the balance equation (\ref{eq:balanceClassical})
we have
\begin{equation}
\int_{\reg}\dot{\rho}\dV+\int_{\bdom}\flux\cdot\nor\dA=\int_{\reg}s\dV.
\end{equation}
Using Gauss's theorem one concludes that 
\begin{equation}
\begin{alignedat}{2}\diver\flux+\dot{\rho} & =s &  & \quad\text{in}\quad\reg,\\
\flux(x)\cdot\nor(x) & =\psi_{\reg}(x) &  & \quad\text{on}\quad\bdom.
\end{alignedat}
\label{eq:diffBalanceGen}
\end{equation}

\begin{rem}
\label{rem: inverseCauchy}Let $\{\psi_{\reg}\}$ be a flux system.
Assume that there is a differentiable vector field $\flux$ such that
for each region $\reg$, $\psi_{\reg}=T\cdot\nor$, then, 
\begin{equation}
\Psi_{\reg}=\int_{\bdom}\psi_{\reg}\dA=\int_{\bdom}\flux\cdot\nor\dA=\int_{\reg}\diver\flux\dV.
\end{equation}
It follows that if $\diver\flux$ is bounded, the boundedness assumption
(\ref{eq:Cauchy-Boundedness}) holds. This observation is an inverse
to Cauchy's theorem.

For detailed, generalized and technical presentations of Cauchy's
flux theory see for example \cite{Schuricht2007} and references cited
therein.
\end{rem}

\subsection{Radiance Systems\label{sub:Radiance-Systems}}

The space of continuous real valued mappings defined on the sphere
will be denoted by $\pdists$. For a regular region $\reg$ in space,
a \emph{radiance} \emph{field} over $\bdry\reg$ is a  mapping 
\begin{equation}
\radi_{\reg}:\bdom\tto\pdists.\label{eq:radianceFieldDefined}
\end{equation}
For $x\in\bdom$ and $\norb\in\sphere$, $\radi_{\reg}(x)(\norb)$,
is interpreted as the density of radiation power per unit area at
$x\in\bdom$ crossing $\bdom$ in the direction $\norb$. It should
be emphasized that this interpretation is given here without any justification
in order that the presentation is adapted to the context of radiation
(see Section \ref{sub:Radiance-Basic-Assumption}).

Clearly, the radiance field may be regarded as a function $\radi'_{\reg}:\bdom\times\sphere\to\reals$
by $\radi'_{\reg}(x,\norb)=\radi_{\reg}(x)(\norb)$. We prefer using
$\radi_{\reg}$ because of the generalization of Section \ref{sec:Modern-Analysis:-Total}.

For a given $\norb\in\sphere$, the \emph{radiance field at the direction}
$\norb$ is the function
\begin{equation}
\radi_{\reg,\norb}:\rthree\tto\reals,\qquad\radi_{\reg,\norb}(x)=\radi_{\reg}(x)(\norb).
\end{equation}
The total emitted power density at $x$, the \emph{irradiance} or
\emph{radiant emittance} is therefore given by
\begin{equation}
\psi_{\reg}(x)=\int_{\sphere}\radi_{\reg}(x)(\norb)\dee\sterad(\norb)=\int_{\sphere}\radi_{\reg}(x)(\norb)\dee\sterad(\norb),
\end{equation}
where $\sterad$ is the solid angle measure. The total power emitted
from $\bdom$ is 
\begin{equation}
\begin{split}P_{\reg}=\int_{\bdom}\psi_{\reg}(x)\dA & =\int_{\bdom}\left(\int_{\sphere}\radi_{\reg}(x)(\norb)\dee\sterad(\norb)\right)\dA.\end{split}
\end{equation}
Fubini's theorem implies that we can write
\begin{equation}
P_{\reg}=\int_{\sphere}\left(\int_{\bdom}\radi_{\reg}(x)(\norb)\dA(x)\right)\dee\sterad=\int_{\sphere}P_{\reg,\norb}\dee\sterad(\norb),
\end{equation}
where 
\begin{equation}
P_{\reg,\norb}=\int_{\bdom}\radi_{\reg}(x)(\norb)\dA(x)=\int_{\bdom}\radi_{\reg,\norb}(x)\dA(x)\label{eq:DirectionalEmissivePower}
\end{equation}
 is the total power emitted from $\reg$ in the direction $u\in\sphere$
(\emph{directional emissive power} in \cite{Mahan2002}). It is noted
that for a fixed $u\in\sphere$, $P_{\reg,\norb}$ is the total flux
of the density $\radi_{\reg,\norb}(x)=\radi_{\reg}(x)(\norb)$---a
scalar flux density over $\bdom$. 

A \emph{radiance system} is a collection $\{\radi_{\reg}\}$ for all
regions $\reg\subset\rthree$. The basic objective of flux theory
is to study the dependence of the radiance $\radi_{\reg}$ on the
region $\reg$. 

We will say that a radiance system $\{\radi_{\reg}\}$ is a \emph{Cauchy
radiance system }if for each $\norb\in\sphere$, the scalar valued
flux system $\{\radi_{\reg,\norb}\}$ satisfies Cauchy's postulates.
From Cauchy's theorem for scalar fluxes, it follows that for a Cauchy
radiance system, for every $\norb\in\sphere$, there is a vector field,
the \emph{radiance vector field for the direction $\norb$,} 

\begin{equation}
\radvec_{\norb}:\rthree\tto\rthree,\quad\text{such that}\quad\radvec_{\norb}(x)\cdot\nor(x)=\radi_{\reg,\norb}(x),\label{eq:ClassCauchyForRad}
\end{equation}
where $\nor(x)$ is the unit normal to $\bdom$ at $x$.

Using the radiance vector field, Equation (\ref{eq:DirectionalEmissivePower})
may be written as 
\begin{equation}
P_{\reg,\norb}=\int_{\bdom}\radvec_{\norb}\cdot\nor\dA.\label{eq:DirectionalPowerWithVectorField}
\end{equation}

The existence of the radiance vector field for the direction $\norb$
enables one to define the field $\radvec$ over $\rthree\times\rthree$,
whose value is a function over the sphere by 
\begin{equation}
\radvec(\nor,x)(\norb)=\radvec_{\norb}(x)\cdot\nor.
\end{equation}
We will refer to $\radvec:\rthree\times\rthree\to\pdists$ as a \emph{radiation
density tensor} and the expression for the power becomes 
\begin{equation}
\begin{split}P_{\reg} & =\int_{\bdom}\left[\int_{\sphere}(\radvec_{\norb}(x)\cdot\nora(x))\,\dee\sterad(\norb)\right]\dee A,\\
 & =\int_{\bdom}\left[\int_{\sphere}\radvec(\nor(x),x)(\norb)\,\dee\sterad(\norb)\right]\dee A,
\end{split}
\label{eq:PowerUsingTensor}
\end{equation}

Clearly, $\radvec$ is linear in $\nor$ and using $\vbase_{j}$ to
denote the $j$-th standard base vector in $\rthree$, one may define
\begin{equation}
\radvec_{j}(x)=\radvec(\vbase_{j},x),
\end{equation}
so that 
\begin{equation}
\radvec(\nor,x)=\sum_{j}\radvec_{j}(x)\nor_{j},
\end{equation}
and
\begin{equation}
\begin{split}P_{\reg} & =\int_{\bdom}\left[\int_{\sphere}\biggl(\sum_{j}\radvec_{j}(x)\nor_{j}\biggr)(\norb)\,\dee\sterad(\norb)\right]\dee A,\\
 & =\int_{\bdom}\sum_{j}\nor_{j}\left[\int_{\sphere}\biggl(\radvec_{j}(x)\biggr)(\norb)\,\dee\sterad(\norb)\right]\dee A,\\
 & =\int_{\bdom}\sum_{j}\nor_{j}\boldsymbol{q}_{j}\dee A,
\end{split}
\end{equation}
where
\begin{equation}
\boldsymbol{q}_{j}=\int_{\sphere}\bigl(\radvec_{j}(x)\bigr)(\norb)\,\dee\sterad(\norb)\label{eq:EnergyFlux-Irrad}
\end{equation}
are the components of the total radiation energy flux vector field
$\boldsymbol{q}$.
\begin{rem}
It is noted that the equations above make it possible to obtain the
total flux across surfaces that are not necessarily the boundaries
of regions. For any smooth surface $S$, one has to replace the integration
over $\bdom$ by integration of $S$.
\end{rem}

\subsection{The Source Term and Differential Balance Equation}

Using Gauss's theorem in Equation (\ref{eq:DirectionalPowerWithVectorField})
one has
\begin{equation}
P_{\reg,\norb}=\int_{\reg}\diver\radvec_{\norb}\dV
\end{equation}
and
\begin{equation}
\begin{split}P_{\reg} & =\int_{\sphere}\left[\int_{\reg}\diver\radvec_{\norb}\dV\right]\dee\sterad,\\
 & =\int_{\reg}\left[\int_{\sphere}\diver\radvec_{\norb}\,\dee\sterad\right]\dee V.
\end{split}
\end{equation}

A balance equation for the energy flux in the direction $\norb\in\sphere$
will be of the form
\begin{equation}
\int_{\reg}\dot{\rho}_{\norb}\dV+\int_{\bdom}\radi_{\reg,\norb}\dA=\int_{\reg}s_{\norb}\dV
\end{equation}
and with Equation (\ref{eq:ClassCauchyForRad})
\begin{equation}
\int_{\reg}\dot{\rho}_{\norb}\dV+\int_{\bdom}\radvec_{\norb}\cdot\nor\dA=\int_{\reg}s_{\norb}\dV.
\end{equation}
The corresponding differential balance equation is
\begin{equation}
\diver\radvec_{\norb}+\dot{\rho}_{\norb}=s_{\norb}.\label{eq:DifferentialBallanceDirectional}
\end{equation}
Integration over the unit sphere will give the balance equation for
the total energy.

\subsection{The Basic Assumption for Radiance\label{sub:Radiance-Basic-Assumption}}

It is noted that thus far very little properties of radiation were
used. One could replace the sphere $\sphere$ by some other space
$B$ so that the space $\pdists$ would be the collection of continuous
functions on $B$. For example, if the space $B$ is the set $\{1,2,3\}$,
the space $U$ of real valued functions on it is identical to $\rthree$
by identifying $\vf_{i}$ with $\vf(i)$ for each $i\in B=\{1,2,3\}$.
Clearly, continuity is insignificant in this example. Such a construction
will lead us to traditional stress theory as we remark below. In fact,
if we do not insist that the functions considered be real valued,
the set $U$ can be any set. We could consider a system $\{K_{\reg,\norb}\}$,
for each region $\reg$ and each $\norb\in B$ and require that it
satisfies Cauchy's postulates for and fixed $\norb$. For each $\norb\in B$,
Cauchy's theorem would imply the existence of a flux vector field
$\radvec_{\norb}$.

Among such theories, what characterizes radiance theory is the assumption
that radiance vector field $\radvec_{u}$ is parallel to $\norb$,
\ie, 
\begin{equation}
\radvec_{\norb}=\radint_{\norb}\norb,\label{eq:BasicAssumptionRadiance}
\end{equation}
where $\radint_{\norb}=\abs{\radvec_{\norb}}$ is a positive real
valued function on $\rthree$. It follows from Equation \eqref{eq:ClassCauchyForRad}
that
\begin{equation}
\radi(x,\nora,\norb)=\radint_{\norb}(x)\norb\cdot\nora.
\end{equation}
Denoting by $\theta$ the angle between the vectors $\norb$ and $\nora$,
\begin{equation}
\radi(x,\nora,\norb)=\radint_{\norb}(x)\cos\theta.
\end{equation}

With this basic assumption, the interpretation of $\radi_{\reg}(x)(\norb)$
as the flux of energy flowing in the direction $\norb$ is justified.
For vectors $\norb$ that point out of $\reg$, $\radi_{\reg}(x)(\norb)$
is the power flux density outgoing from $\reg$ and for vectors $\norb$
that point into $\reg$, $\radi_{\reg}(x)(\norb)$ is the incoming
power density. One can also consider surfaces that need not be the
boundaries of regions so that ``incoming'' and ``outgoing'' are
meaningless. In such circumstances $\radi_{\reg}(x)(\norb)$ and $\radi_{\reg}(x)(-\norb)$
are specified independently in order to distinguish between fields
such as those generated in the middle between two equal light sources
and fields where there are no light sources.

Equation (\ref{eq:DirectionalPowerWithVectorField}) may now be written
as
\begin{equation}
P_{\reg,\norb}=\int_{\bdom}\radint_{\norb}\norb\cdot\nor\dA.\label{eq:DirectionalPowerWithAssumption}
\end{equation}
Thus, the expression \eqref{eq:PowerUsingTensor} for the power becomes
\begin{equation}
\begin{split}P_{\reg} & =\int_{\bdom}\left[\int_{\sphere}\radint_{\norb}(x)\cos\theta(\norb)\,\dee\sterad(\norb)\right]\dee A,\\
 & =\int_{\bdom}\left[\int_{\sphere}\radint_{\norb}(x)\norb\,\dee\sterad(\norb)\right]\cdot\nor(x)\,\dee A,
\end{split}
\label{eq:PowerWithBasicAssump}
\end{equation}
so that the radiation energy flux vector field is given by
\begin{equation}
\boldsymbol{q}(x)=\int_{\sphere}\radint_{\norb}(x)\norb\,\dee\sterad(\norb).\label{eq:TotalFluxQ}
\end{equation}
Alternatively, we may write
\begin{equation}
\begin{split}P_{\reg} & =\int_{\sphere}\left[\int_{\bdom}\radint_{\norb}(x)\nor\,\dee A\right]\cdot\norb\,\dee\sterad(\norb),\\
 & =\int_{\sphere}\boldsymbol{S}_{\norb}\cdot\norb\,\dee\sterad,
\end{split}
\end{equation}
where 
\begin{equation}
\boldsymbol{S}_{\norb}=\int_{\bdom}\radint_{\norb}(x)\nor\,\dee A\label{eq:DefRadiantIntensityDirected}
\end{equation}
represents the total contribution of the radiation in the direction
of $\norb$ (\emph{cf.} radiant intensity $\mathscr{I}$ in \cite[Equation (2.1.9)]{Apresyan1996}.

Comparing Equation (\ref{eq:PowerWithBasicAssump}) with traditional
expositions of radiation theory, \eg,  \cite{Planck,Chandrasekhar,Apresyan1996,Modest,SobolevRadiation},
it is noted that $\radint_{\norb}(x)$ is the scalar radiance density,
\eg,  $K(x,\norb),$ that they use.
\begin{rem}
\label{CompareStresses}In order to get some insight to the various
objects defined and derived here, we compare the notions under consideration
to stress theory of continuum mechanics. In continuum mechanics, the
traction vector $\sfc$, determined by its 3 components $\sfc_{i}$,
depends on the point $x$ and external unit normal $\nora$. Thus,
while $\sfc$ is determined by the 3 components $t_{i}$ the radiation
field is determined but the infinite number of ``components'' $\radi_{\reg,\norb}(x)=\radi_{\reg}(x)(\norb)$
 parametrized by the direction $\norb$.

The vector field $\radvec_{\norb}$ that induces the radiance, is
thus analogous to the $i$-th row of the stress tensor $\st$, $\st_{i}$,
that gives the $i$-th component of the traction vector by
\begin{equation}
t_{i}=\st_{i}\cdot\nora.
\end{equation}
Again, just as there are 3 such rows in the stress tensor, there are
uncountably infinite row vectors $\radvec_{\norb}$, each for any
value of $\norb$.

The property of radiation implying that the radiance flux vector for
the direction $\norb$ is parallel to $\norb$ is therefore analogous
to the property that the $i$-th row of the stress tensor is parallel
to the $i$-th coordinate axis. Thus, the analogous property for stresses
is that the stress matrix is diagonal.
\end{rem}

We may now apply Green's theorem to Equation (\ref{eq:DefRadiantIntensityDirected})
to obtain
\begin{equation}
\boldsymbol{S}_{\norb}=\int_{\reg}\nabla\radint_{\norb}\dV.
\end{equation}
Thus, the total flux out of $\reg$ may be expressed as 
\begin{equation}
\begin{split}P_{\reg} & =\int_{\sphere}\left[\int_{\reg}\nabla\radint_{\norb}\dV\right]\cdot\norb\,\dee\sterad,\\
 & =\int_{\reg}\left[\int_{\sphere}\nabla\radint_{\norb}\cdot\norb\,\dee\sterad\right]\dee V.
\end{split}
\end{equation}
This also follows from 
\[
\diver\radvec_{\norb}=\diver(\radint_{\norb}\norb)=\nabla\radint_{\norb}\cdot\norb.
\]
Evidently, the same result could be obtained by applying Green's theorem
to the second line of Equation (\ref{eq:PowerWithBasicAssump}). The
differential balance equation for the radiation in the direction $\norb$,
Equation (\ref{eq:DifferentialBallanceDirectional}), assumes the
form
\begin{equation}
\nabla\radint_{\norb}\cdot\norb+\dot{\rho}_{\norb}=s_{\norb}.\label{eq:DifferentialBallanceDirectionalWithAssump}
\end{equation}

\begin{rem}
\label{ConservationOfRadiance} Assume that $\dot{\rho}_{\norb}=s_{\norb}=0$
in the last equation. If follows that
\begin{equation}
\nabla\radint_{\norb}\cdot\norb=0,
\end{equation}
\ie,  the directional derivative in the direction $\norb$ of the
radiance in the direction $\norb$ vanishes. This result, formulated
as the conservation of $i_{\norb}$ under translation in the direction
$\norb$, is sometimes referred to as the \emph{fundamental theorem
of radiometry} (\eg, \cite[pp. 18--19]{Boyd1983}).
\end{rem}

For the differential balance equation (\ref{eq:diffBalanceGen}) corresponding
to the total radiation energy 
\begin{equation}
\diver\boldsymbol{q}+\dot{\rho}=s,
\end{equation}
one has
\begin{equation}
\diver\boldsymbol{q}=\int_{\sphere}\nabla\radint_{\norb}\cdot\norb\,\dee\sterad.
\end{equation}

\begin{rem}
From Remark \ref{rem: inverseCauchy}, it follows that if one starts
from the standard expression (\ref{eq:PowerWithBasicAssump}), then,
Cauchy postulates follow for differentiable fields $\radint_{\norb}$.
In other words, the Cauchy postulates are both necessary and sufficient
for the traditional radiation expression. Not merely some, but all
differentiable radiance fields represent Cauchy radiance systems.
Specifically, it follows from Equation (\ref{eq:DirectionalPowerWithAssumption})
that
\begin{equation}
P_{\reg,\norb}=\int_{\reg}\diver\!(\radint_{\norb}\norb)\dV
\end{equation}
and so
\begin{equation}
\abs{P_{\reg,\norb}}\les\max_{x\in\rthree}\abs{\diver\!(\radint_{\norb}\norb)(x)}\abs{\reg},
\end{equation}
the boundedness expression for $P_{\reg,\norb}$.
\end{rem}

\subsection{Virtual Power}

Let $\norb$ be a fixed direction is space and let $\vf_{\norb}:\bdom\to\reals$
be a field. One may consider the action
\begin{equation}
P_{\reg,\norb}(\vf_{\norb})=\int_{\bdom}\radi_{\reg,\norb}(x)\vf_{\norb}(x)\,\dee A(x).
\end{equation}
We interpret $P_{\reg,\norb}(\vf_{\norb})$ as follows. Let $W_{\norb}:\bdom\to\reals^{+}$
be a differentiable function that we interpret as a distribution of
radiance meters over the boundary that measure the radiance in the
direction $\norb$ so that $W_{\norb}(x)$ is the density of the meters
at $x$. The field $\vf_{\norb}$ is conceived as a variation of $W_{\norb}$.
Thus, while $W_{\norb}(x)$ should be positive by our interpretation,
this limitation does not apply to $\vf_{\norb}(x)$. Hence, $P_{\reg,\norb}(\vf_{\norb})$,
is interpreted as the change of the total power of radiation propagating
in the direction $\norb$ measured by the distribution of meters under
the variation $\vf_{\norb}$ of the distribution. For short, we will
refer to $P_{\reg,\norb}(\vf_{\norb})$ as the \emph{virtual power}.

Using the radiance vector field $\radvec_{\norb}$ we may write the
virtual power as
\begin{equation}
P_{\reg,\norb}(\vf_{\norb})=\int_{\bdom}\vf_{\norb}(x)\radvec_{\norb}(x)\cdot\nor(x)\,\dee A(x).
\end{equation}
The last equation may be transformed using Gauss's theorem into
\begin{equation}
\begin{split}P_{\reg,\norb}(\vf_{\norb}) & =\int_{\reg}\diver\!(\vf_{\norb}\radvec_{\norb})\,\dee V,\\
 & =\int_{\reg}\nabla\vf_{\norb}\cdot\radvec_{\norb}\dV+\int_{\reg}\vf_{\norb}\diver\radvec_{\norb}\dV,
\end{split}
\end{equation}
and with Equation (\ref{eq:DifferentialBallanceDirectional}) we obtain
\begin{equation}
\int_{\bdom}\radi_{\reg,\norb}\vf_{\norb}\,\dee A+\int_{\reg}\vf_{\norb}(\dot{\rho}_{\norb}-s_{\norb})\dee V=\int_{\reg}\nabla\vf_{\norb}\cdot\radvec_{\norb}\dV.
\end{equation}
Using the basic assumption for radiance (\ref{eq:BasicAssumptionRadiance})
we finally arrive at
\begin{equation}
\int_{\bdom}\radi_{\reg,\norb}\vf_{\norb}\,\dee A+\int_{\reg}\vf_{\norb}(\dot{\rho}_{\norb}-s_{\norb})\dee V=\int_{\reg}\radint_{\norb}\nabla\vf_{\norb}\cdot\norb\dV,
\end{equation}
where evidently $\nabla\vf_{\norb}\cdot\norb$ is the directional
derivative of $\vf_{\norb}$ in the direction of $\norb$.

\section{Measure Valued Radiance\label{sec:Modern-Analysis:-Total}}

\subsection{Preliminaries}

In the foregoing analysis, the flux density of radiation energy is
distributed continuously on the sphere $\sphere$. For example, for
a point $x\in\bdom$, a function $K_{x}=\radi_{\reg}(x):\sphere\to\reals$
describes the distribution of radiation power, where for $\norb\in\sphere$,
$K_{x}(\norb)$ was interpreted as the radiation density emitted in
the direction of $\norb$. It is noted that if one wishes to consider
power radiated in one particular direction $\norb_{0}$, a distribution
such as $K_{x}$ above is not general enough. One has to include mathematical
objects such as the Dirac delta distribution, $\delta_{\norb_{0}}$,
representing the emission of power in the particular direction $\norb_{0}$
and which is a measure on the sphere. It is therefore assumed henceforth
that the distribution of radiation, or the radiance,\emph{ }at $x\in\bdom$,
is a Borel measure on the unit sphere. Thus, if we denote the radiance
at $x$ by $\tradi_{x}$, for any Borel subset $D$ of the sphere
$\sphere$, $\tradi_{x}(D)$ denotes the measure of $D$ indicating
the total power density per unit area emitted through the pencil at
$x$ subtended by $D$. In particular, the irradiance is given by
\begin{equation}
E=\tradi_{x}(\sphere)=\int_{\sphere}\dee\tradi_{x}.
\end{equation}
In the traditional continuous case, the measure $\tradi_{x}$ is absolutely
continuous relative to the solid angle measure, or area measure on
the unit sphere, so that there is a Radon-Nikodym derivative
\begin{equation}
K_{x}=\frac{\dee\tradi_{x}}{\dee\sterad},\qquad\radi_{x}:\sphere\tto\reals
\end{equation}
and 
\begin{equation}
\tradi_{x}(D)=\int_{D}\radi_{x}\dee\sterad.
\end{equation}
(See for example \cite{Preisendorfer1965,ArvoThesis95} who postulate
the continuity requirement right from the start.) As before, the function
$\radi_{x}(\norb)$ indicates the radiance at $x$ out of $\reg$
in the direction of $\norb$ and the total density of power, emitted
at $x$ in all directions is 
\begin{equation}
\int_{\sphere}\radi_{x}\dee\sterad.
\end{equation}

We will denote by $\measrs$ the vector space of Borel measures on
the unit sphere, and so, $\tradi_{x}\in\measrs$ (in comparison with
$\radi_{x}\in\pdists$). The vector space operations are naturally
defined in $\measrs$ by $(a\tradi_{x}+a'\tradi_{x}')(D)=a\tradi_{x}(D)+a'\tradi'_{x}(D)$.
In addition, one can define a norm on $\measrs$ providing it with
the structure of a Banach space.

It is recalled that the space of Borel measures $\measrs$ may be
identified with the dual space $\pdists^{*}$ of the space $\pdists$
of continuous real valued mappings on the sphere. Specifically, the
measure $\mradi_{x}$ may be regarded as a continuous and linear functional
$\mradi'_{x}:\pdists\to\reals$ such that ''
\begin{equation}
\mradi'_{x}(\pmd)=\int_{\sphere}\pmd(\norb)\dee\mradi_{x}(\norb).
\end{equation}
The expression above is interpreted as the virtual power density at
$x$ for the change of distribution of radiance meters $\pmd$ at
$x$. In the sequel, we will not distinguish in the notation between
the measure and the linear functional it induces. In terms of this
action, a natural norm on $\measrs$ is given by 
\begin{equation}
\norm{\mradi_{x}}=\sup_{\pmd\in\pdists}\frac{\mradi_{x}(\pmd)}{\norm{\pmd}},
\end{equation}
where $\norm{\pmd}=\max_{\norb\in\sphere}\abs{\pmd(\norb)}$.

\subsection{Measure Valued Radiance Systems and Radiance Tensors}

In analogy with the definition of the radiance field over $\bdom$
in (\ref{eq:radianceFieldDefined}), the basic assumption is that
for every region $\reg\subset\rthree$ there is a smooth function,
the \emph{measure valued radiance}, 
\begin{equation}
\tradi_{\reg}:\bdom\tto\measrs.\label{eq:RadSystemMeasrs}
\end{equation}
The collection $\{\mradi_{\reg}\}$, for all regions $\reg$ is a
\emph{measure valued radiance system}. 

For a given measure valued radiance system and an element $\pmd\in\pdists$,
let $\{\mradi_{\reg,v}\}$ be the flux system such that 
\begin{equation}
\mradi_{\reg,\pmd}(x)=\mradi_{\reg}(x)(\pmd)=\int_{\sphere}v\,\dee(\mradi_{\reg}(x)).
\end{equation}
\ie,  the action of $\mradi_{\reg}(x)$ on $\pmd$ is the action
of a continuous linear functional on a continuous function defined
on the sphere. Similarly, one can evaluate the integral of the measure
$\mradi_{\reg}(x)$ over any integrable Borel subset $D\subset\sphere$
to yield
\begin{equation}
\mradi_{\reg}(x)(D)=\int_{D}\dee(\mradi_{\reg}(x)).
\end{equation}

In analogy with Section \ref{sub:Radiance-Systems} we now assume
that for each fixed $\pmd\in\pdists$, the measure valued radiance
systems $\{\mradi_{\reg,\pmd}\}$ satisfies Cauchy's postulates. Let
$\pmd\in\pdists$ be a fixed distribution. It follows from Cauchy's
theorem for scalar valued flux systems that there is a vector field
$\mrad_{\pmd}=\sum_{i}\mrad_{\pmd i}\vbase_{i}:\rthree\to\rthree$
such that 
\begin{equation}
\mradi_{\reg,\pmd}(x)=\mradi_{\reg}(x)(\pmd)=\mrad_{\pmd}(x)\cdot\nor(x)=\sum_{i}\mrad_{\pmd i}(x)\nor_{i}(x).\label{eq:CauchyForMeasuresBasic}
\end{equation}
for any $x\in\bdom$. It follows from Equation (\ref{eq:CauchyForMeasuresBasic})
that $\mrad_{\pmd}$ and $\mrad_{\pmd i}$ depend on $\pmd\in\pdists$
continuously and linearly. As $\mrad_{\pmd i}(x)$ are real, there
are three elements $\mrad_{i}(x)\in\measrs$ such that $\mrad_{\pmd i}(x)=\mrad_{i}(x)(\pmd)$.
Hence,
\begin{equation}
\mradi_{\reg,\pmd}(x)=\sum_{i}\nor_{i}(x)\mrad_{i}(x)(\pmd)=\sum_{i}\nor_{i}(x)\int_{\sphere}\pmd\,\dee(\mrad_{i}(x)).
\end{equation}
Clearly, $\mrad_{i}(x)\in\measrs$ are the three flux distributions
at $x$ corresponding to the area elements which are perpendicular
to the standard base vectors in $\rthree$. 

Using the notation $L(\rthree,\measrs)$ for the space of linear mappings
$\rthree\to\measrs$ (which are also continuous as they are defined
on a finite dimensional space), one can now define the mapping
\begin{equation}
\mrad:\rthree\tto L(\rthree,\measrs)\cong\measrs\otimes\rthree
\end{equation}
by
\begin{equation}
(\mrad(x)(\nor))(\pmd)=\mrad_{\pmd}(x)\cdot\nor=\sum_{i}\nor_{i}(x)\mrad_{i}(x)(\pmd).\label{eq:Define-I}
\end{equation}
Observe that for any $\nor\in\rthree$, 
\begin{equation}
\begin{split}\sum_{i}\mrad_{i}(x)\nor_{i} & =\mrad(x)(\nor),\\
 & =\mrad(x)\biggl(\sum_{i}\nor_{i}\vbase_{i}\biggr),\\
 & =\sum_{i}\nor_{i}\mrad(x)(\vbase_{i}),
\end{split}
\end{equation}
so that $\mrad_{i}(x)=\mrad(x)(\vbase_{i})\in\measrs$. Thus, we may
introduce the notation
\begin{equation}
\mrad=\sum_{i}\mrad_{i}\otimes\vbase_{i},\label{eq:ComponentsOfMeasRadiation}
\end{equation}
and write
\begin{equation}
\mradi_{\reg}=\mrad\cdot\nor=\biggl(\sum_{i}\mrad_{i}\otimes\vbase_{i}\biggr)\cdot(\nor_{j}\vbase_{j})=\sum_{i}\mrad_{i}\nor_{i}=\mrad(\nor).\label{eq:CauchyMeasuresWithTensorI}
\end{equation}
The infinite dimensional tensor $\mrad$ will be referred to as the
\emph{measure valued radiance tensor}. In particular, for any Borel
integrable set $D$,
\begin{equation}
\begin{split}\mradi_{\reg}(x)(D) & =\sum_{i}\mrad_{i}(x)(D)\nor_{i}(x)=\mrad(x)(D)\cdot\nor(x),\\
 & =\sum_{i}\nor_{i}(x)\int_{D}\dee(\mrad_{i}(x)).
\end{split}
\end{equation}

In the case where the measures $\mrad(x)$ may be represented by densities
relative to the solid angle measure $\sterad$ on the sphere, one
has
\begin{equation}
\dee\mrad(x)=\radvec(\nor,x)\dee\sterad
\end{equation}
where $\radvec(\nor,x)$ is the radiance density tensor.

Using the measure valued radiance tensor, the total energy flux out
of a region $\reg$ is given as 
\begin{equation}
\begin{split}P_{\reg} & =\int_{\bdom}\left[\int_{\sphere}\nor(x)\cdot\dee(\mrad(x))(\norb)\right]\dA(x),\\
 & =\int_{\bdom}\left[\int_{\sphere}\dee(\mrad(x))(\norb)\right]\cdot\nor(x)\dA(x),\\
 & =\int_{\bdom}\mrad(x)(\sphere)\cdot\nor(x)\dA(x),
\end{split}
\end{equation}
where, 
\begin{equation}
\mrad(x)(\sphere)=\sum_{i}\mrad_{i}(x)(\sphere)\otimes\vbase_{i}
\end{equation}
and $\mrad_{i}(x)(\sphere)$, $i=1,2,3$, represent the total measures
of the sphere. The energy flux vector field $\bs q$ is therefore
given by
\begin{equation}
\bs q(x)=\int_{\sphere}\dee(\mrad(x))(\norb)=\mrad(x)(\sphere).
\end{equation}

\subsection{Measure Valued Radiance Tensors: Totals Approach}

We present below an alternative derivation of the measure valued radiance
tensor in terms of a total radiation distribution corresponding to
a region $\reg$. In order to formulate the theory, one needs to use
integration and differentiation of functions on $\rthree$ valued
in a Banach space---the space $\measrs$ in our case. (See for example
\cite[Chap. V]{Lang2} or \cite[Chap. VIII]{Dieudonne1960} for some
details on calculus of functions valued in Banach spaces.)

\subsubsection{The Analogs of Cauchy's Postulates}

As in the definition of the radiance field over $\bdom$ in (\ref{eq:RadSystemMeasrs}),
the basic assumption is that for every region $\reg\subset\rthree$
there is a smooth function, the measure valued radiance, $\tradi_{\reg}:\bdom\tto\measrs.$
The \emph{total} of the radiation corresponding to the region $\reg$
is defined as the distribution
\begin{equation}
\total_{\reg}=\int_{\bdom}\tradi_{\reg}\dA\in\measrs,
\end{equation}
where we integrate the measures $\tradi_{\reg}(x)$ over $\bdom$.
In other words, for each measurable subset $D\subset\sphere$, $\Phi_{\reg}(D)$
is the total flux of energy flowing in the directions included in
the pencils determined by $D$ from all points in $\bdom$.

In case $\total_{\reg}$ is absolutely continuous relative to the
solid angle measure on the sphere, the directional emissive power
$P_{\reg,\norb}$ is given by the Radon-Nikodym derivative
\begin{equation}
P_{\reg,\norb}=\frac{\dee\total_{\reg}}{\dee\sterad}(\norb).
\end{equation}

The following assumptions are made in analogy with Section \ref{sub:Traditional-Cauchy-Fluxes}.

\subsubsection*{Locality:}

It is assumed that for a point $x\in\bdom$, $\tradi_{\reg}(x)$ depends
on $\reg$ through the outwards pointing normal $\nor(x)$ to $\bdry\reg$
at $x$.

It follows that there is a function
\begin{equation}
\tradi:\rthree\times S^{2}\tto\measrs,
\end{equation}
to which we will refer as the Cauchy mapping, such that
\begin{equation}
\tradi(x,\nor(x))=\tradi_{\reg}(x)
\end{equation}
for any region $\reg$ such that $x\in\bdom$ and $\nor(x)$ is the
outwards pointing normal to $\reg$ at $x$.

Using the Cauchy mapping we may rewrite the total as 
\begin{equation}
\total_{\reg}=\int_{\bdom}\tradi(x,\nor(x))\dA.\label{eq:total-cauchy}
\end{equation}

\subsubsection*{Boundedness (Balance):}

We assume that there is a positive number $C$ such that 
\begin{equation}
\norm{\total_{\reg}}\les C\abs{\reg},\label{eq:boundedness-t}
\end{equation}
where $\norm{\Phi_{\reg}}$ is the norm in $\measrs$ of the total.

\subsubsection*{Regularity:}

It is assumed that the Cauchy mapping $\tradi$ is smooth.

\subsubsection{The Analog of Cauchy's Theorem}

We outline a sketch of the proof of Cauchy's theorem for the current
settings. Consider an infinitesimal tetrahedron containing the point
$x\in\rthree$ such that for $\ga=0,1,2,3$, $A_{\ga}$ are the areas
of the faces of the tetrahedron and $\nor_{\ga}$ are the unit normals
to the faces. Thus, Equation (\ref{eq:total-cauchy}) assumes the
form 
\begin{equation}
\total_{\reg}=\sum_{\alpha}\tradi(x,\nor_{\alpha})A_{\alpha},
\end{equation}
and the boundedness assumption (\ref{eq:boundedness-t}) implies that
\begin{equation}
\norm{\total_{\reg}}=\biggl\Vert\sum_{\alpha}\tradi(x,\nor_{\alpha})A_{\alpha}\biggr\Vert\les C\abs{\reg}.
\end{equation}
Since $\sum_{\alpha}\nora_{\alpha}A_{\alpha}=0$, we have 
\begin{equation}
\nora_{0}=-\sum_{p=1}^{3}\frac{A_{p}}{A_{0}}\nora_{p},
\end{equation}
which implies
\begin{equation}
\left\Vert \tradi\biggl(x,-\sum_{p=1}^{3}\frac{A_{p}}{A_{0}}\nor_{p}\biggr)+\sum_{p=1}^{3}\tradi(x,\nor_{p})\frac{A_{p}}{A_{0}}\right\Vert \les C\frac{\abs{\reg}}{A_{0}}.
\end{equation}
As the size of the tetrahedron approaches zero, the right-hand side
of the last equation tends to zero. Taking the limit, it follows that
the left-hand side of the last equation vanishes. Since for any norm
on $\measrs$, $\norm{J_{0}}=0$ implies that $\tradi_{0}=0\in\measrs$,
we obtain
\begin{equation}
\tradi\left(x,-\sum_{p=1}^{3}\frac{A_{p}}{A_{0}}\nor_{p}\right)=-\sum_{p=1}^{3}\frac{A_{p}}{A_{0}}\tradi(x,\nor_{p}).
\end{equation}
The last equation implies that the dependence of $\tradi$ on its
second argument $\nor$ is linear. 

We conclude that there is a field 
\begin{equation}
\mrad:\rthree\tto L(\rthree,\measrs)\qquad\text{given by}\qquad\mrad(x)(\nor)=\tradi(x,\nor).
\end{equation}
Using the mapping $\mrad$, one can write the analog of Cauchy's formula
as
\begin{equation}
\mrad(x)(\nor)=\tradi_{\reg}(x)\label{eq:CauchyFormula-t}
\end{equation}
for every region $\reg$ such that $\nor$ is the outwards pointing
normal to $\bdom$ at $x$. Hence, $\mrad$ is identical to the measure
valued radiance tensor of (\ref{eq:Define-I}).

Using the measure valued radiance tensor, the irradiance, the total
of the radiation may be written as
\begin{equation}
\total_{\reg}=\int_{\bdom}\mrad(x)\cdot\nor(x)\dA\in\measrs.\label{eq:total-with-tensor}
\end{equation}
Using the Gauss theorem for the measure valued field $\mrad$ we may
rewrite Equation (\ref{eq:total-with-tensor}) as
\begin{equation}
\Phi_{\reg}=\int_{\reg}\diver\mrad\dV.
\end{equation}

The measure $\total_{\reg}$ may be integrated over the sphere (or
any other Borel measurable subset thereof) to give the total power
\begin{equation}
P_{\reg}=\int_{\sphere}\dee\Phi_{\reg}=\int_{S^{2}}\dee\left(\int_{\bdom}\mrad(\nor)(x)\dA\right).
\end{equation}

\subsection{The Basic Assumption of Radiation for Measure-Valued Radiation Tensors}

The basic assumption for radiance theory should be generalized so
that it applies to the measure valued radiance tensor. The assumption
cannot be formulated as in (\ref{eq:BasicAssumptionRadiance}) because
in the general case one cannot assume that the vector density $\radvec_{u}$
exists. Nevertheless, the basic constitutive assumption may be generalized
as follows.

We recall that a measure $\mu$, viewed as a continuous, linear functional
acting on continuous functions, may be multiplied by an integrable
function $\phi$ to yield the measure $\phi\odot\mu$ defined by 
\begin{equation}
(\phi\odot\mu)(f)=\mu(\phi f),\text{\,\,\,\ or equivalently,\,\,\ }\int f\,\dee(\phi\odot\mu)=\int f\phi\,\dee\mu.
\end{equation}
In addition, the Radon-Nykodim theorem implies that if there is a
positive measure $\lambda$ such that $\lambda(D)=0$ implies that
$\mu(D)=0$, then there is an integrable function, the Radon-Nykodim
derivative 
\begin{equation}
f=\frac{\dee\mu}{\dee\lambda},
\end{equation}
 such that $\mu=f\odot\lambda$.

We now use a procedure as in \cite[pp. 236--239]{Edwards1965}. For
each $x\in\rthree$ which is kept fixed in this paragraph, consider
the positive measure
\begin{equation}
\abs{\mrad(x)}\equiv\abs{\mrad}(x):=\biggl[\sum_{j}\mrad_{j}^{2}(x)\biggr]^{\shalf},\quad\text{where,}\quad\left[\mrad_{j}^{2}(x)\right](D)=\left[\mrad_{j}(x)(D)\right]^{2}.
\end{equation}
Clearly, for each $j=1,2,3$, $\abs{\mrad_{j}(x)(D)}\les\abs{\mrad(x)(D)}$
for every measurable subset $D$. It follows from the Radon-Nikodym
theorem that for each $j$ there is an integrable function $\wh{\mrad}_{j}(x)$
defined on $\sphere$ such that 
\begin{equation}
\mrad_{j}(x)=\wh{\mrad}_{j}(x)\odot\abs{\mrad(x)}\quad\text{and}\quad\sum_{j}\wh{\mrad}_{j}^{2}(x)(\norb)=1
\end{equation}
for almost all $\norb\in\sphere$ relative to the measure $\abs{\mrad(x)}$.
For example, for any measurable subset $D$,
\begin{equation}
\int_{D}\dee\left(\mrad_{j}(x)\right)=\int_{D}\wh{\mrad}_{j}(x)\dee\abs{\mrad(x)}.
\end{equation}

One may define the vector function $\wh{\mrad}(x):\sphere\to\rthree$
on the sphere by 
\begin{equation}
\wh{\mrad}(x)(\norb)=\sum_{j}\wh{\mrad}_{j}(x)(\norb)\vbase_{j}
\end{equation}
and so
\begin{equation}
\mrad(x)=\sum_{j}\mrad_{j}(x)\vbase_{j}=\wh{\mrad}(x)\odot\abs{\mrad(x)},
\end{equation}
where $\odot$ in the equation above indicates the product of the
measure $\abs{\mrad(x)}$ by the vector valued integrable function
$\wh{\mrad}(x)$. In other words, all the singularities of the measures
$\mrad_{j}(x)$ are included in $\abs{\mrad(x)}$ and the ``unit
vector'' $\wh{\mrad}(x)$ distributes them to the various directions.

The basic assumption of radiation theory may be formulated simply
as
\begin{equation}
\wh{\mrad}(x)(\norb)=\norb.\label{eq:basic-Assump-Meas}
\end{equation}
In other words, for each $x\in\rthree$ there is a scalar measure
$\abs{\mrad(x)}$ such that 
\begin{equation}
\mrad(x)=\incl\odot\abs{\mrad(x)},\label{eq:basic-Assump-Meas-sec}
\end{equation}
where 
\begin{equation}
\incl:\sphere\to\sphere
\end{equation}
is the identity mapping on the sphere. Again, $\odot$ on the right-hand
side of Equation (\ref{eq:basic-Assump-Meas-sec}) is the product
of a vector (or a sphere)-valued function and a measure.

It follows from Equations (\ref{eq:Define-I}--\ref{eq:ComponentsOfMeasRadiation})
that
\begin{equation}
\tradi(x,\nor)=\mrad(x)\cdot\nor=\abs{\mrad(x)}\odot(\norb\cdot\nor).\label{eq:CauchyMeasuresWithAssump}
\end{equation}
Using Equation (\ref{eq:CauchyMeasuresWithAssump}) the total distribution
may be written now as 
\begin{equation}
\Phi_{\reg}=\int_{\bdom}\abs{\mrad}\odot(\norb\cdot\nor)\dA,
\end{equation}
and using Gauss's theorem
\begin{equation}
\Phi_{\reg}=\int_{\reg}\sum_{j}\frac{\partial\abs{\mrad}}{\bdry x_{j}}\odot\norb_{j}\dV=\int_{\reg}(\nabla\abs{\mrad})\odot\norb\dV,
\end{equation}
where the gradient of the measure valued function $\abs{\mrad}$ is
used as well as the notation $(\nabla\abs{\mrad})\odot\norb=\sum_{j}\frac{\partial\abs{\mrad}}{\bdry x_{j}}\odot\norb_{j}$.

For any measurable $D\subset\sphere$, 
\begin{equation}
\begin{split}\int_{D}\dee(\mrad(x)\cdot\nor) & =\int_{D}\dee(\abs{\mrad(x)}\norb\cdot\nor),\\
 & =\int_{D}\norb\cdot\nor\,\dee\abs{\mrad(x)}.
\end{split}
\end{equation}
In the classical case where $\dee\abs{\mrad(x)}=\radint_{\norb}(x)\dee\sterad$
we revert to 
\begin{equation}
\int_{D}\dee(\mrad(x)\cdot\nor)=\int_{D}\norb\cdot\nor\,\radint_{\norb}(x)\dee\sterad.
\end{equation}

\end{document}